\documentclass{aa}

\usepackage{amsmath}
\usepackage{latexsym}
\usepackage{txfonts}
\usepackage{graphicx}
\usepackage{natbib}
\usepackage[latin1]{inputenc}
\usepackage{nicefrac}
\usepackage{color}
\usepackage{ulem}

\begin{document}
\normalem

\title{Investigation of mass flows in the transition region and corona in a three-dimensional numerical model approach}

\author{P. Zacharias \inst{1}\and H. Peter \inst{2}\and S. Bingert\inst{2}}
\institute{Kiepenheuer-Institut f\"ur Sonnenphysik, Sch\"oneckstrasse 6,
  79104 Freiburg, Germany \and 
Max Planck Institute for Solar System Research, Max-Planck Strasse 2, 37191 Katlenburg-Lindau, Germany}

\abstract%
{
The origin of solar transition region redshifts is not completely understood. Current research is addressing this issue by investigating three-dimensional magneto-hydrodynamic models that extend from the photosphere to the corona. 
}{
By studying the average properties of emission line profiles synthesized from the simulation runs and comparing them to observations with present-day instrumentation, we investigate the origin of mass flows in the solar transition region and corona. 
}{
Doppler shifts were determined from the emission line profiles of various extreme-ultraviolet emission lines formed in the range of $T=10^4-10^6$ K. Plasma velocities and mass flows were investigated for their contribution to the observed Doppler shifts in the model. In particular, the temporal evolution of plasma flows along the magnetic field lines was analyzed. 
}{
Comparing observed vs. modeled Doppler shifts shows a good correlation in the temperature range $\log(T$/[K])=4.5-5.7, which is the basis of our search for the origin of the line shifts.
The vertical velocity obtained when weighting the velocity by the density squared is shown to be almost identical to the corresponding Doppler shift. Therefore, a direct comparison between Doppler shifts and the model parameters is allowed. A simple interpretation of Doppler shifts in terms of mass flux leads to overestimating the mass flux. 
Upflows in the model appear in the form of cool pockets of gas that heat up slowly as they rise. Their low temperature means that these pockets are not observed as blueshifts in the transition region and coronal lines.
For a set of magnetic field lines, two different flow phases could be identified. The coronal part of the field line is intermittently connected to subjacent layers of either strong or weak heating, leading either to mass flows into the loop (observed as a blueshift) or to the draining of the loop (observed as a redshift).    
}
{
}

\titlerunning{Mass flows in the transition region}
\maketitle


\section{Introduction} 
Persistent average redshifts of emission lines from the solar transition region were observed by \cite{doschek:1976} for the first time. Early observations from rockets (e.g., Hassler et al., \citeyear{hassler:1991}) had limited potential and did not provide a conclusive picture of the nature of these redshifts. The evaluation of SUMER (Solar Ultraviolet Measurements of Emitted Radiation, Wilhelm et al., \citeyear{wilhelm:1995}) observations onboard SOHO led to the conclusion that redshifts in the transition region predominately present vertical motions that turn into blueshifts at higher temperatures (Peter, \citeyear{peter:1999a}; Peter \& Judge, \citeyear{peter+judge:1999}). These redshifts have been explained by a variety of models including the return of previously heated spicular material (Athay \& Holzer, \citeyear{athay+holzer:1982}; Athay, \citeyear{athay:1984}) or nanoflare-induced waves generated in the corona that would produce a net redshift in transition region lines (Hansteen, \citeyear{hansteen:1993}). For a more comprehensive discussion of coronal loops in relation to the Doppler shift observations see \cite{peter+judge:1999}.
Peter et al. (\citeyear{peter+al:2004}, \citeyear{peter+al:2006}) found that three-dimensional numerical models spanning the region from the photosphere to the corona are producing redshifts in transition region lines.\\
In these models, coronal heating comes from the Joule dissipation of currents produced by the braiding of the magnetic field through photospheric plasma motions; however, details on the origin of the redshifts have not been identified so far. \cite{spadaro:2006} used one-dimensional coronal loop models to show that the redshifts can be the result of transient heating near the loop footpoints. Recently, \cite{hansteen:2010} has shown that transition region redshifts are naturally produced in episodically heated models where the average volumetric heating scale height lies between 50 and 200 km. The authors suggest that chromospheric material is heated in place to coronal temperatures or that rapidly heated coronal material leads to a conduction front heating the chromospheric material below. Both options lead to a high-pressure plug of material at upper transition region temperatures that relaxes towards equilibrium by expelling material.

In this paper, we describe the investigation of a small active region by means of a three-dimensional (3D) magneto-hydrodynamic (MHD) model. We synthesize transition region lines that  have been typically observed by SUMER. The synthesized emission may be compared directly to observations. The mass balance in the model simulation is investigated as is the relation of the Doppler shifts to the magnetic structure in order to elucidate the nature of the transition region redshifts.

The 3D MHD model is introduced in section \ref{model}. Section \ref{results} focuses on the spectral line synthesis from the simulation results and on the comparison with observations. The origin and nature of the modeled Doppler shifts is investigated in section \ref{understanding}, before results are discussed in section \ref{discussion}. 

\section{The coronal 3D MHD model}\label{model}

We are applying a 3D MHD model of the solar atmosphere above a small active region in a computational box of the size 50x50x30 Mm$^3$. A detailed description of the model can be found in \cite{bingert:2010}. 
The simulations are carried out using the Pencil-code \citep{brandenburg+dobler:2002}, a high-order finite-difference code for compressible MHD. The set of equations being solved includes the continuity equation, the induction equation, the Navier-Stokes equation, and the energy equation with an anisotropic Spitzer heat conduction term \citep{spitzer:1962} and an optical thin radiative loss function \citep{cook:1989}.

For the lower boundary we constructed a map of the magnetic field based on two observed magnetograms, one from an active region, spatially scaled down by a factor of five, and another one from a quiet Sun area. Thus, the resulting magnetic map represents the quiet Sun with some activity added by a pore and is supposed to be typical for the magnetic field underlying the transition region of the Sun outside active regions.

The main difference between the model presented in this study and the one described by \cite{bingert:2010} is the higher grid point resolution of our simulation, allowing for a slightly lower magnetic diffusivity. The resolution of $256^3$ grid points corresponds to a grid spacing of 195 km in the horizontal direction and 75-285 km in the vertical direction, where a nonequidistant grid is applied.
This corresponds to a maximum resolvable vertical temperature gradient of about 2\,K/m, when assuming a temperature difference of $10^6$\,K across five grid points. This number is confirmed by the 3D model discussed by \cite{peter+al:2006}, see their Fig.\,5. In a static 1D empirical model based on an emission measure inversion, \cite{mariska:1992} found a maximum temperature gradient of the order of 10\,K/m (his Table 6.1).
Flows present in our 3D model are assumed to smooth out steep temperature gradients, resulting in temperature changes that are less substantial than expected for the case of the static 1D empirical models \citep[see Fig.\,5 of][]{peter+al:2006}. Thus, it can be assumed that temperature gradients in the 3D model are being properly resolved. 
However, to synthesize transition region emission lines, the temperatures and densities as computed in the 3D MHD model have to be interpolated \citep[Sect.\,\ref{synthesis}; for details see][]{peter+al:2006}.

The viscosity and heat conduction terms as published by \cite{spitzer:1962} are applied; i.e., in the coronal part of the box, the dynamic viscosity is set to $\nu{\approx}10^{11}\,{\rm{m}}^2{\rm{s}}^{-1}$ and the thermal diffusivity is about $\chi{\approx}10^{12}\,{\rm{m}}^2{\rm{s}}^{-1}$. For velocities of up to 100 km/s and a grid scale as used in our simulation, the average Reynolds number Re and the average Prandtl number Pr are of the order of 0.1.
The magnetic diffusivity is chosen in a way that the average magnetic Reynolds number Rm is near unity, i.e., $\eta{\approx}10^{10}\,{\rm{m}}^2{\rm{s}}^{-1}$. This ensures that the current sheets that are being formed (i.e., the gradients in magnetic field) can still be resolved in the computation.
This choice of parameters implies that the average magnetic Prandtl number is ${\rm{Pm}}{=}\nu/\eta{\approx}10$, i.e., on average, magnetic field gradients are larger than velocity field gradients.
The above-mentioned \emph{average} dimensionless numbers can only be used for a rough characterization, because they vary considerably by orders of magnitude in the computational domain, depending on the local plasma conditions.

While the average magnetic Prandtl number is greater than unity, implying that the velocity field is more effective in amplifying field on the smallest scales than the magnetic field, still the dissipation of energy in the corona is dominated by Ohmic heating processes, and viscous heating only plays a minor role. The deposited energy is efficiently redistributed by heat conduction, the rate of which is comparable to the Ohmic heating rate, except for low temperatures, where radiative losses dominate.

\begin{figure}
\begin{center}
\includegraphics[width=0.8\columnwidth]{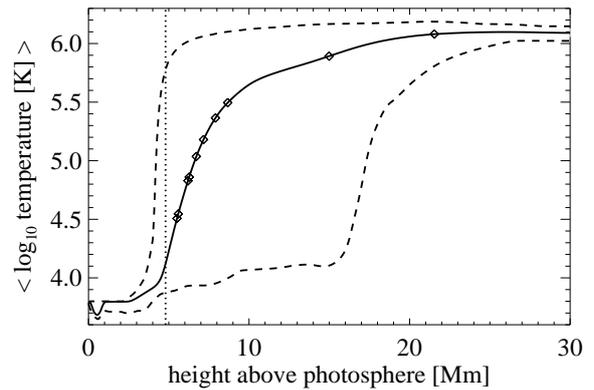}
\caption{Height dependence of the average temperature in horizontal layers of the simulation box (solid line). Diamonds indicate the formation temperature of the emission lines that are investigated in this study (compare Table \ref{formation_temperature}, column labeled $\log T_{contrib}$). The dashed lines show the minimum and maximum temperature in the horizontal layers.}  
\label{logT_mean}
\end{center}
\end{figure}

The  boundary conditions are chosen such that the simulation box is periodic in (x,y) and closed for mass flux on both the top and bottom. At the bottom boundary, the temperatures and densities are set to the average photospheric temperature and density values at $\tau{=}1$. The magnetic field at the lower boundary is driven by horizontal motions mimicking the photospheric horizontal granular motions. This leads to the braiding of the magnetic field lines and the induction of currents, which are subsequently dissipated by Ohmic heating. As a result, the plasma is heated and driven, leading to a dynamic corona. The simulation was run - for some 30 minutes - to a state where the behavior is independent of the initial condition. This point of time is defined as $t=0$ min in our study. The simulation produces a dynamic transition region and corona with flows up to and above the sound speed and with the horizontally {\emph{averaged}} quantities, such as the average temperature (see Fig.\,\ref{logT_mean}) and density as a function of height, showing only a little variation over time. 
Below about 4\,Mm, the computational domain contains material at and below chromospheric temperatures. This height encompasses ten chromospheric scale heights, i.e., a drop in density by more than four orders of magnitude.
This dense part of the box acts as a mass reservoir for evaporating plasma into the corona and for material draining from above. It also serves as a reservoir for energy that is conducted back from the corona without being radiated in the transition region.

The energy deposition rate in the corona is highly variable in space and time. 
A reasonable estimate of $Q{\approx}10^{-5}\,{\rm{W}}\,{\rm{m}}^{-3}$ for the energy deposition at a height of some 10\,Mm corresponds to an energy flux density of some $100\,{\rm{W}}\,{\rm{m}}^{-2}$ at the base of the corona \citep[see][their Fig.\,5]{bingert:2010}.
A typical length scale on which current sheets are forming is $\ell{\approx}1$\,Mm. Thus, the associated time scale for magnetic dissipation is of the order of $\tau{=}\ell^2/\eta{\approx}100$\,s.
Therefore, the deposited energy density can be estimated to be $E{=}Q\tau{\approx}10^{-3}\,{\rm{J}}\,{\rm{m}}^{-3}$.
According to the magnetic energy density $B^2/(2\mu_0)$, this corresponds to a field strength of about $B{\approx}5$\,G.
This number is within an order of magnitude of the actual magnetic field strength in the coronal part of the computational domain, which is about 10\,G at 10\,Mm height.
Consequently, the field line braiding in our model can lead to an accumulation of magnetic energy within an order of magnitude in energy to the corresponding potential field extrapolated from the photosphere. If the driving in the photosphere is switched off, the heating will only continue for the duration of the dissipation time $\tau$, i.e., some minutes.
Thus, there is a need for a continuous energy supply through the photospheric motions.
The above-mentioned order of magnitude estimations are confirmed by a more detailed analysis of the numerical experiment.

\section{Results}\label{results}
\subsection{Spectral line synthesis}\label{synthesis}
Optically thin transition region and coronal emission lines are synthesized from the results of the MHD calculations (Table \ref{formation_temperature}).  
In the low-density plasma of the solar transition region and corona, spontaneous radiative decay is the main process that depopulates excited levels. The rate of this process is given by $A_{ji}\cdot n_j$, where $n_j$ denotes the number density of atoms in the excited state $j$, and $A_{ji}$ is the Einstein coefficient for a spontaneous transition from level $j$ to $i$. The total emissivity of the transition is
\begin{equation}
\epsilon\, (T,n_e) = h\nu_{ji}\cdot A_{ji}\cdot n_j = G(T) \cdot n_e^2, \label{emiss}
\end{equation}
where $n_e$ is the electron number density and $G(T)$ the temperature dependent contribution function
\begin{equation}
G(T)=h\nu_{ji}\cdot A_{ji}\cdot \frac{n_j}{n_e \cdot n_i}\cdot \frac{n_i}{n_{ion}}\cdot \frac{n_{ion}}{n_{el}}\cdot \frac{n_{el}}{n_H}\cdot \frac{n_H}{n_e},
\end{equation}
that comprises the relative abundance of the ionic species, the abundance of the element relative to hydrogen, and the number density of hydrogen relative to that of the electrons. These terms are accessible through atomic data bases, such as CHIANTI (Dere et al., \citeyear{dere+al:1997}; Landi et al., \citeyear{landi+al:2006}), and can be easily evaluated once the temperatures and densities are known. A rather strong assumption that has been used in these calculations is to assume ionization equilibrium. However, \cite{peter+al:2006} showed that ionization equilibrium is a fairly good approximation that is valid for most parts in the simulation box.

Nonetheless, it must be noted that future numerical experiments at higher resolution capable of resolving steeper temperature and velocity gradients will have to overcome this simplification. This has already been done in 1D models, but is not yet feasible in 3D computations.

The MHD calculations were performed using a nonequidistant grid in the vertical direction with a spacing of neighboring grid points down to 75 km in the transition region, but the temperature resolution in these data is still insufficient for calculating the line emissivities in the transition region. To reach a temperature resolution below the typical contribution function widths of transition region lines ($\log (T /\mbox{[K]})\approx$ 0.2-0.3), the MHD results are interpolated in the vertical direction. The vertical distance between grid points in the transition region is thus reduced down to 40 km in the transition region and temperature gradients of 0.1 in $\log T$ per grid cell are obtained on average at each height. All parameters discussed in the following were derived from such spatially interpolated data.

The Doppler shift of the respective line (cf. Table \ref{formation_temperature}) is derived by calculating the first moment of the line profile. The average line shift is obtained by averaging the line profiles horizontally before integrating along the line-of-sight.

Table \ref{formation_temperature} shows the wavelengths and formation temperatures of the lines synthesized in this study. The lines are chosen so as to cover the entire transition region in temperature, ranging from $T\approx 30.000$ K to $T\approx 1.3\times 10^6$ K. Also, these lines can be observed with instruments like SUMER/SOHO (e.g., Peter \& Judge, \citeyear{peter+judge:1999}). Temperatures labeled $\log T_{line}$ are the temperatures, where the term $\frac{n_{ion}}{n_{el}}T^{-1/2}\exp \left(-\frac{h\nu}{k_BT}\right)$, i.e., the product of maximum ionization fraction and excitation rate, of the respective line has its maximum. The temperatures labeled $\log T_{contrib}$ indicate the temperatures, where the maximum contribution is found in the simulation box (i.e., bulk of the emission in the respective line originates from regions with this temperature). The good correspondence between $\log T_{line}$ and $\log T_{contrib}$ reflects that the transition region is basically isobaric in the simulation.

\begin{table}
\begin{center}
\caption{Emission lines synthesized in this study together with their wavelengths and line formation temperatures.\label{formation_temperature}}
\begin{tabular}{lrcc}
\hline
\hline
line & wavelength & \multicolumn{2}{c}{line formation temperature}\\
     & [\AA]        & $\log ( T_{line} / [\mbox{K}] )$
                    & $\log ( T_{contrib} / [\mbox{K}] )$
                    \\
\hline
Si\,{\sc{ii}}\tablefootmark{(1)} &  1533.4 & 4.44 & 4.5\\
C\,{\sc{ii}}\tablefootmark{(1)}  &  1334.5 & 4.59 & 4.6\\
Si\,{\sc{iv}} &  1393.8 & 4.88 & 4.9\\
C\,{\sc{iii}} &   977.0 & 4.91 & 4.8\\
C\,{\sc{iv}}  &  1548.2 & 5.02 & 5.0\\
O\,{\sc{iv}}  &  1401.2 & 5.19 & 5.2\\
O\,{\sc{v}}   &   629.7 & 5.40 & 5.4\\
O\,{\sc{vi}}  &  1031.9 & 5.48 & 5.5\\
Ne\,{\sc{viii}} & 770.4 & 5.80 & 5.9\\
Mg\,{\sc{x}}  &   624.9 & 6.05 & 6.1\\
\hline
\end{tabular}
\tablefoot{
\tablefoottext{1}{The formation temperatures for Si\,{\sc{ii}} and C\,{\sc{ii}} as listed here are most likely overestimated (see end of Sect.\,\ref{synthesis})}
}
\end{center}
\end{table}

Owing to the strong heating at the footpoints of the corona and the resulting strong increase of density towards the chromosphere, cool lines like Si\,{\sc{ii}} and C\,{\sc{ii}} tend to be formed well below their ionization equilibrium temperature. This issue is discussed in detail by \cite{judge:2003} and \cite{peter+al:2006}.
For consistency with the previous investigations outlined in this paper, e.g., when plotting Doppler shift vs. line formation temperature, we refer to the line formation temperatures listed in Table \ref{formation_temperature} for all ions, including Si\,{\sc{ii}} and C\,{\sc{ii}}. 
 
%


\subsection{Doppler shifts derived from the MHD model}
\begin{figure}
\begin{center}
\includegraphics[width=0.8\columnwidth]{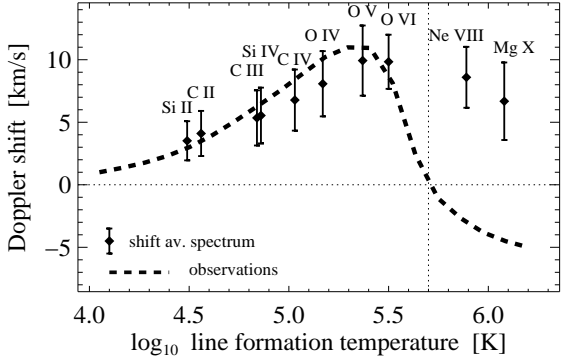}
\caption{Average Doppler shifts derived from MHD model as function of line formation temperature. The height of the bars indicates the width of the distribution obtained by averaging horizontally over all line profiles at each height in the simulation box.}  
\label{temp_doppler}
\end{center}
\end{figure}

Average Doppler shifts for each emission line of Table \ref{formation_temperature} are calculated from the emission line profiles at each time step. The line profile functions at each grid point are assumed to be Gaussian shaped and of the form 
\begin{equation} 
I_{v}=\frac{I_{tot}}{\sqrt{\pi} \omega_{th}} \exp \left[-\frac{(v-v_{los})^2}{\omega_{th}^2}\right],
\end{equation}
where
\begin{equation}
\omega_{th} = \left( \frac{2k_BT}{m}\right)^2
\end{equation}
is the thermal width, $I_{tot}$ is the emissivity (Eq. \ref{emiss}) at the respective grid point, and the vertical direction is assumed to be the line-of-sight. To obtain the Doppler shift of each emission line, the line profiles of all grid points are averaged horizontally at each height, and the average line profile is integrated along the line-of-sight. The first moment of this function corresponds to the shift of the average spectrum, i.e., the net Doppler shift of the respective emission line (cf. Fig. \ref{temp_doppler}).

The investigation of these average Doppler shifts shows that, in contrast to earlier work (e.g., Peter et al., \citeyear{peter+al:2004}; \citeyear{peter+al:2006}), a much better representation of the variation in average redshifts with line formation temperature is obtained for the simulation run of our model setup up to $\log (T$/[K])$\approx$5.6, i.e., the low corona. Figure \ref{temp_doppler} shows a comparison between observed (data from Peter \& Judge, \citeyear{peter+judge:1999}) and modeled Doppler shifts that show a close correlation at transition region temperatures up to $\log (T$/[K])=5.6.
Due to the high spatial resolution of the simulation, the dynamics in the simulation box are better resolved, especially in the transition region where the temperature gradients are large.
Also, the setup of the numerical experiment is more representative of the Sun outside of active regions (for which the transition region redshifts are well documented) than it was in earlier studies by, e.g., \cite{gudiksen+nordlund:2005a} (see section 2).  
However, the observed blueshifts in the coronal part at $\log (T$/[K])$>$ 5.7 are not reproduced correctly by the MHD model owing to the impenetrable upper boundary of the simulation box. 
As discussed by \cite{peter+al:2004}, this upper boundary condition is expected to have no influence on the transition region redshifts produced in the magnetically closed structures in the computational domain, but might prevent the formation of blueshifts in the coronal lines in the framework of our numerical experiment. On the real Sun, magnetically open regions might be connected to the solar wind and thus be responsible for the blueshifts of coronal lines. However, the upper boundary condition of the model prevents the outflows and thus could suppress the observed blueshifts of coronal lines.
%


\section{Understanding transition region Doppler shifts}\label{understanding}
The compliance between the observed transition region redshifts and the average Doppler shifts in the MHD model (cf. Fig. \ref{temp_doppler}) allows the investigation of the origin of these line shifts based on our MHD model.

\subsection{Coronal mass balance}\label{balance}
\begin{figure}
\begin{center}
\includegraphics[width=0.8\columnwidth]{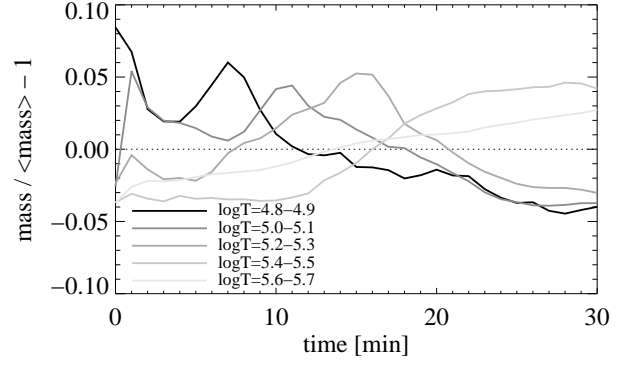}
\caption{Temporal evolution of the relative change of mass in a given temperature interval as indicated in the legend. The average mass over time in the respective temperature interval is denoted by $<\textrm{mass}>$.
}  
\label{mass}
\end{center}
\end{figure}

As a first approach to tackling the line shifts in the model, the temporal variation in the total mass, i.e., the overall number of particles, in various temperature intervals in the simulation box is determined. 
The temperature interval is chosen to be $d \log (T$/[K])=0.1. Figure \ref{mass} shows examples for the variation in the mass in temperature intervals ranging from $\log (T$/[K])=4.8-4.9 to $\log (T$/[K])=5.6-5.7. The mass enclosed fluctuates between 2\% and 5\% during the 30-minute simulation time.

In principle, this small change in mass (or number of particles $N$) within a temperature interval could be due to an average down- (or up-)flow, which could cause the observed average downflow. However, 
an estimate of the average vertical velocity based on the observed variation in the number of particles $\delta = (N-\langle N \rangle)/\langle N \rangle \approx$ 5\%,  the transition region scale height $H{\approx}5$\,Mm, and the timescale of $\tau{\approx}15$\,min implies $\langle v \rangle= \delta{\cdot}N/(n A \tau) = \delta{\cdot}H/\tau \approx 0.3$ km/s, a value much too low to explain the observed redshifts.%
\footnote{$N = n A H$ is the total number of particles in the corona, $n = 10^{15}$ m$^{-3}$ is the particle density in the transition region, and $A$ is a measure for the area corresponding to a horizonal cut through the box.}

\begin{figure}
\begin{center}
\includegraphics[width=0.8\columnwidth]{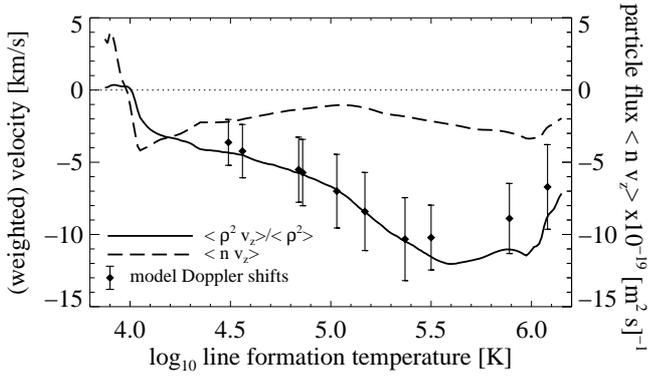}
\caption{
Weighted velocity and mass flux as function of temperature in the simulation box.
The solid line shows the intensity-weighted velocity (Sect.\,\ref{balance}). This matches the Doppler shifts from the synthesized spectra (bars, for comparison, taken from Fig.\,\ref{temp_doppler}). The average mass flux in the vertical direction is shown as a dashed line.
}
\label{weighting}
\end{center}
\end{figure}

Because the intensity of optically thin emission lines scales with the density squared, the Doppler shifts can be approximated by $\langle \rho^2  v_z\rangle / \langle \rho^2 \rangle$, where $\langle\dots\rangle$ represents the average in a small temperature interval. This 
{\emph{intensity-weighted}} velocity, of course, reflects the trend of the average Doppler shifts with line formation temperature, as illustrated in Fig.\,\ref{weighting}.

In Fig. \ref{weighting}, the velocities and densities that are evaluated are confined to regions in the simulation box within the respective temperature interval and above a height of $z=4.8$ Mm. The minimum temperature at this height is $\log (T/$[K])=3.8. It is assumed that plasma with temperatures below this threshold corresponds to chromospheric material (cf. Fig. \ref{logT_mean}) rather than to transition region and coronal material. Since the purpose of this study is to investigate only the velocities of the transition region and coronal plasma, chromospheric temperature areas are not considered. Otherwise, the material on average would be at rest, and a zero net mass flux would be observed. For temperatures above $\log (T$/[K])=6.0, the small number
of data points in the simulation box would result in unreliable predictions of the
model, which is why such temperatures are not considered in the analysis.

%
Cool material can reach high altitudes through pressure imbalances or local dynamics, as shown by \cite{peter+al:2006}, e.g., in their Fig 1$\gamma,\delta$. These upwelling cool plasma fingers are not a good representation of spicules, because they lack the fine structure and dynamics of observed spicules. These numerical experiments are not realistic enough to produce spicules, nonetheless they show the elevation of cool material into the corona.

\subsection{Pockets of cool gas}
A close inspection of Fig. \ref{weighting} shows an upward directed intensity-weighted velocity at temperatures below $\log (T$/[K])=4.0.
More important, the particle flux $\langle n\,v_z\rangle$ shows a change in sign at $T{\approx}10^4$\,K (dotted line in Fig.\,\ref{weighting}). 
The {\emph{upwards}} directed mass flux at low temperatures ($T<10^4$\,K) is comparable in magnitude to the {\emph{downwards}} directed mass flux at temperatures above $T=10^4$\,K. This would suggest that cool material is supplying the mass into the corona, where it is heated and then rains down. This is further supported by the finding that upflows ($v_z > 0$) are identified up to heights of 10 Mm and 15 Mm for temperatures of $\log (T$/[K]) $\le$ 4.0 and $\log (T$/[K]) $\le$ 5.0 in the simulation box.\\
These findings indicate a process by which cool pockets of material are pushed upwards from the chromosphere into the corona, while being heated slowly.  
The cool pockets are not visible at transition-region and coronal temperatures, since the heating occurs on a somewhat slower timescale than the rise of the material. Therefore, they cannot be observed as blueshifts in transition region and coronal lines. 
This appears to be a phenomenon similar to the idea that cool spicular material is actually supplying mass to the corona that is raining down after it has been heated, as proposed, e.g., by \cite{athay+holzer:1982}.

\subsection{High-pressure plug of material that relaxes towards equilibrium by expelling material} 
\cite{hansteen:2010} suggest that the blueshifts in the low corona ($\log (T$/[K]) $>$ 5.6) can be explained as a natural consequence of episodically heated models, where the heating per particle is concentrated in the upper chromosphere, transition region, or the lower corona. When a heating event occurs in the upper chromosphere or transition region, cooler material is heated rapidly to coronal temperatures. In the lower corona, thermal conduction increases rapidly with rising coronal temperature, and in the case of a heating event, the chromospheric material below can be heated by a thermal wave. If these processes happen expeditiously, the plasma will be heated more or less ``in place'' before it has time to move. Thus, newly heated transition material will be characterized by pressures higher than hydrostatic, and the high-pressure material will expand into regions of lower pressure on either side: downwards towards the chromosphere and upwards towards the corona.

We have found evidence for such a process in one of our simulation runs although on a larger scale. Plasma is ejected as a consequence of increased heating and high pressure in the lower transition region. Results are going to be presented in a second publication. This scenario as proposed by \cite{hansteen:2010} or earlier by \cite{spadaro:2006} is quite different from the one discussed above in Sect. 4.3. Probably, both processes occur in parallel, i.e., upwards accelerated cool parcels of plasma, as well as rapidly heated material in the low atmosphere. Future studies will have to show which, if any, of these two processes is responsible for the persistent Doppler shifts in the transition region.

\subsection{Field line related mechanism causing upflows and downflows}
So far, only vertical flows of material in the simulation domain have been investigated (Sect.\,\ref{balance}). The following discussion will focus on plasma flows \emph{along} magnetic field lines. 

In principle, magnetic field lines can be traced easily in the simulation domain by following the magnetic field vector at a given grid point for a given distance. However, it is not straightforward to follow such a magnetic field line over time, and some assumptions have to be made in order to develop a suitable technique. 
It is assumed that the field lines and the plasma are ``frozen-in'' due to the high electric conductivity of the plasma in the solar atmosphere. 
Thus, the velocity vector of the plasma at the corresponding
time step and site (e.g., the top of the field lines) can be used to determine the displacement
of the starting points of the field lines for the next time step. The highest
point of each field line is used, because it is assumed that the field lines are more widely separated
from each other the higher they are up in the atmosphere. Thus, errors resulting from jumping
between different field lines can be minimized. The magnetic field vector is tracked
in three dimensions from the starting point until the field line returns to the bottom
or leaves the simulation box at the top. Field lines that leave the simulation box
on either side can still be tracked as periodic boundary conditions are applied in the
horizontal direction.
The field line tracking is started in the regions of strong blueshift ($v_D < -8$ km/s)
and in the layer of highest C\,{\sc{iv}} (1548 \AA) emission in the simulation box.

The finite electric conductivity in our numerical experiment means that working with ``frozen-in'' field lines is only an approximation.  
The velocity of the slippage between magnetic field and plasma can be expressed by the term $(\eta/\tau)^{1/2}$, and is of the order of 1\,km/s, calculated on the basis of the magnetic diffusivity $\eta{\approx}10^{10}\,{\rm{m}}^2{\rm{s}}^{-1}$ and the time interval $\tau{\approx}30$\,min over which the field line is being followed.
The calculated velocity of 1\,km/s is below the velocity of the ``frozen-in'' field lines (perpendicular to the magnetic field), and therefore the above assumption is justified.

\begin{figure}
\begin{center}
\includegraphics[width=\columnwidth]{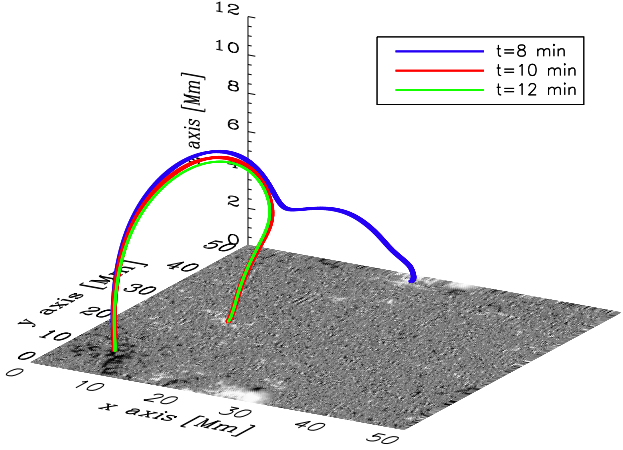}
\caption{Temporal evolution of a single, selected magnetic field line near X-type reconnection site for four minutes simulation time. In addition, the vertical magnetic field component $B_z$ scaled from -1000 G (black) to +1000 G (white) is shown at height zero ($z=0$).}  
\label{fieldline}
\end{center}
\end{figure}

\begin{figure}
\begin{center}
\includegraphics[width=0.8\columnwidth]{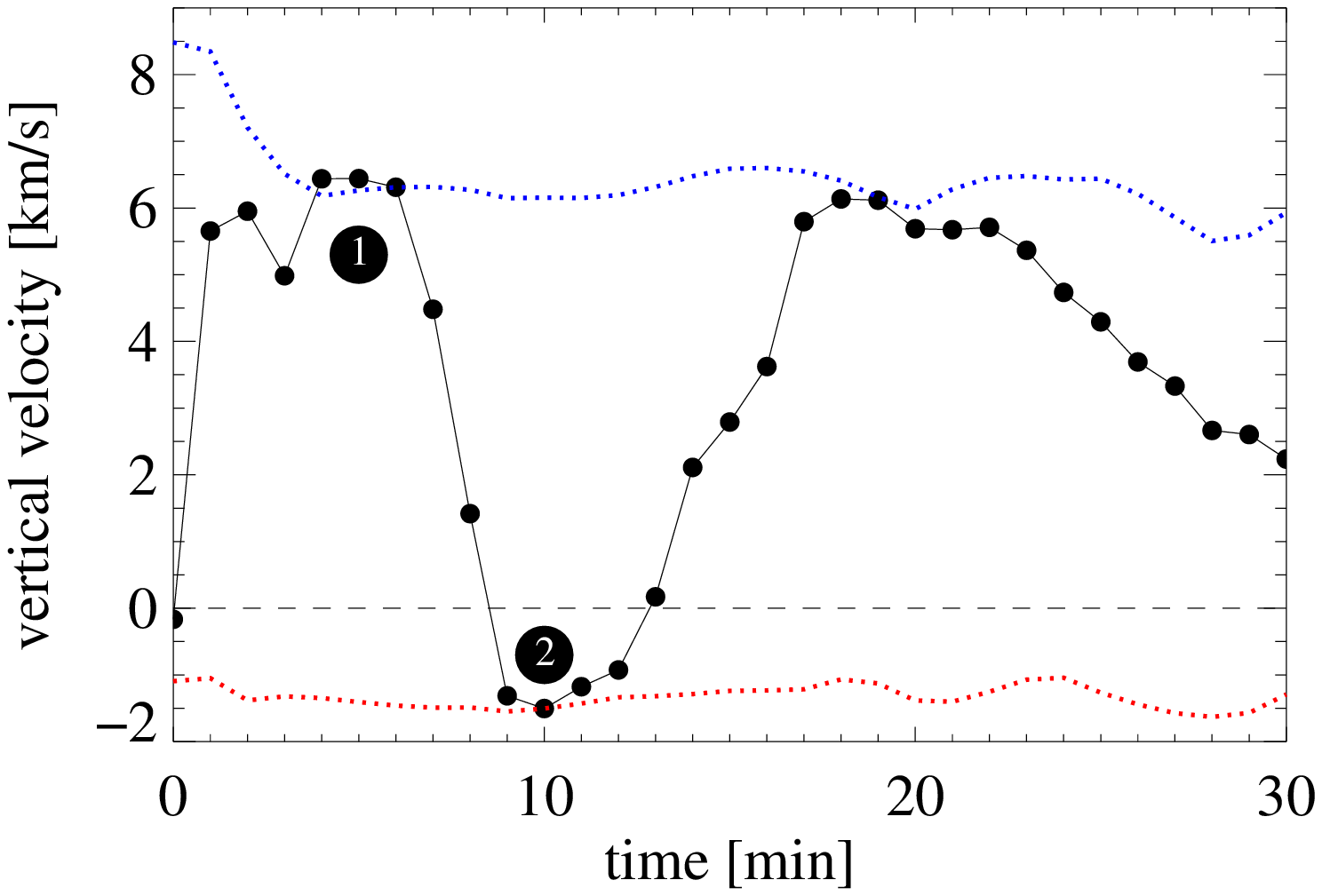}
\caption{Temporal evolution of the vertical velocity $v_z$ along the field line followed in Fig. \ref{fieldline} at the position of highest C\,{\sc{iv}} (1548 \AA) intensity. (1) and (2) indicate the two different field line configurations shown in Fig. \ref{fieldline}, associated with different flow patterns along the field line. (1): Material is flowing upwards along the field line at the site of highest C\,{\sc{iv}} (1548 \AA) corresponding to an observed blueshift. (2): Material is flowing downwards along the field line at the site of highest C\,{\sc{iv}} intensity corresponding to an observed redshift.}  
\label{vz_plot}
\vspace{0.3cm}

\includegraphics[width=0.8\columnwidth]{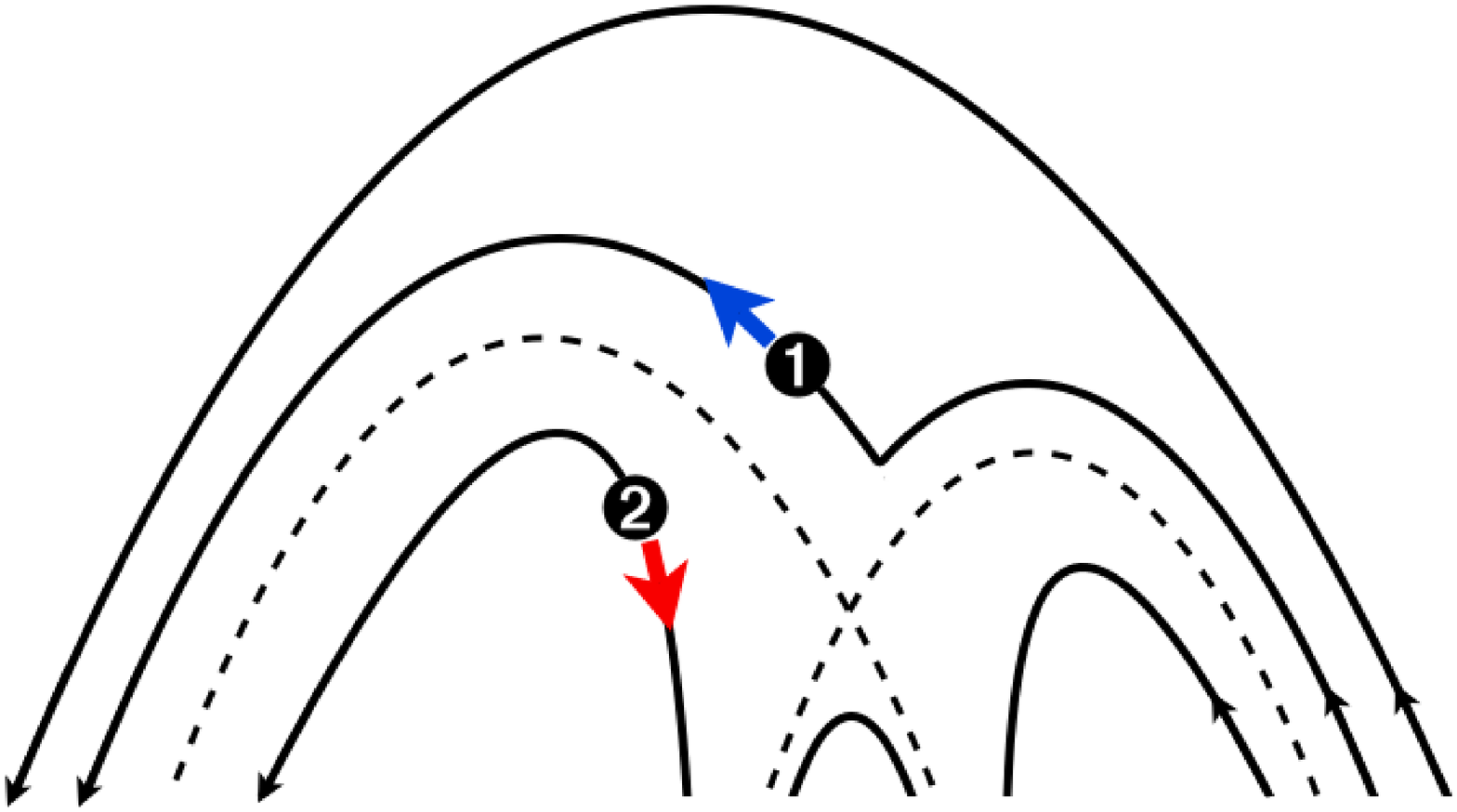}
\caption{Two-dimensional illustration of the changing magnetic field line configuration shown in Fig. \ref{fieldline}. The phases labeled 1 and 2 correspond to phases named identically in Fig. \ref{vz_plot}.}  
\label{sketch}
\end{center}
\end{figure}

Field lines near a small X-type reconnection site have been selected for further analysis. 
The temporal evolution of a single field line is shown in a 3D view in Fig. \ref{fieldline} for four minutes (t=8-12 min) of simulation time. 
The vertical velocity $v_z$ along the magnetic field line is determined at each time step, since this is what the Doppler shift would be showing if the box is viewed from the top (situation at disk center). In Fig. \ref{vz_plot}, $v_z$ is shown at the location of highest C\,{\sc{iv}} (1548 \AA) intensity along the field line (since this is the location representing the observable Doppler shift). A quantitative analysis of the velocities shows a change in sign as the field line reconnects (from blue to red, Fig. \ref{fieldline}). In a first phase (t=0-8 min), a positive velocity is observed at the location of highest intensity along the line corresponding to an upflow of material. Whenever the field line reconnects, the vertical velocity switches signs.  
Plasma is draining out of the loop during the period of time when the field line is connected to a region different than in the first phase. Another connectivity change in the field line occuring in the time period between t=12 min and t=13 min is not shown in Fig. \ref{fieldline}. The vertical velocity switches signs to positive again (cf. Fig. \ref{vz_plot}), corresponding to an upflow of material within the loop. Vertical velocities of two spatially fixed regions of mass loading ($v_z > 0$ at t=6 min) and draining ($v_z < 0$ at t=10 min) are indicated by the red and blue dotted lines in Fig. \ref{vz_plot}. The temporal evolution of these velocities shows little variation, indicating that these sites remain rather stable during the simulation.

Figure \ref{sketch} shows a schematic view of the field line configuration during the upflow (indicated by 1, blue) and downflow (indicated by 2, red) phases. The phases labeled 1 and 2 correspond to phases named identically in Fig. \ref{vz_plot} and are used to illustrate the two-step process outlined above. The coronal part of the magnetic field line is initially connected to a region below, where the material is being heated and accelerated. This leads to an upflow of material into the corona that is being observed as a blueshift along the loop (phase 1). After magnetic reconnection has led to a reconfiguration of the field, the magnetic field line is connected to a different chromospheric region with less heating from below. As a consequence, the material flows back down along the loop leading to the observed redshift (phase 2). A net redshift is observed, since the upflows and downflows take place at different temperatures and densities (cf. section \ref{balance}).

\section{Discussion and conclusions}\label{discussion}
This study of the origin of Doppler shifts in a 3D MHD model extending from the photosphere to the corona reveals persistent redshifts at transition region temperatures. The comparison of the modeled to the observed Doppler shifts shows a close similarity between both quantities with large apparent downflows at transition region temperatures. 
The relation between the Doppler shifts and the mass flux in the model can be analyzed as a next step. 
The synthesized Doppler shifts are reproduced well by an ``intensity weighted'' velocity, i.e., the velocity weighted by the density squared, $\langle{\rho^2}\,v\rangle / \langle{\rho^2}\rangle$. 
The basis for this is that the emissivity is proportional to the density squared ($\propto \rho^2$).

The coronal mass is approximately constant in these simulations, varying between 0.2\% and 0.5\% on average. This variation only cannot explain the persistent transition region Doppler shifts that are a few km/s. This discrepancy is mitigated by the finding that a simple interpretation of the Doppler shifts in terms of an average vertical mass flux leads to a systematic overestimation of the mass flux, an effect that is most obvious in the upper transition region.
The presence of upflows at temperatures below $\log(T$/[K])=4.0 indicates a process by which cool pockets of material are pushed upwards into the corona from below. As they rise, they heat up slowly. Due to their low temperature, they are not observable as blueshifts in the transition region and coronal lines.

When investigating the magnetic field topology, there is evidence for a scenario in which magnetic field lines are intermittently connected to subjacent regions where heating takes place on different scales. This leads to either upflows observed as blueshifts or downflows observed as redshifts along the field lines. The regions of mass loading and draining are fairly stable in space and time and can thus serve as mass reservoirs or sinks. Plasma on field lines traversing these regions experiences either an upflow or a downflow depending on the connectivity of the field line.

As a conclusion, the good correlation between observed and model-synthesized transition region Doppler shifts below ${\log}(T/$[K])${=}5.6$ is due to cool pockets of plasma propelled upwards and heated to transition region and coronal temperatures afterwards, a process similar to the one previously suggested by, e.g., \cite{athay+holzer:1982}. This is in contrast to recent work by Hansteen et al. (2010), who suggest that chromospheric material is heated fast in the low atmosphere and expands into the corona afterwards. Future investigations will have to decide on the merits of these alternatives.

{
\acknowledgements

We thank the referee, Phil Judge, for very useful comments on the manuscript.
This research has been supported by the Deutsche Forschungsgemeinschaft (DFG).
}

\bibliography{literature}
\bibliographystyle{aa}

\end{document}